\documentclass[preprint,review,12pt,authoryear]{elsarticle}

\usepackage{amsmath,amssymb}

\usepackage{booktabs}
\usepackage{multirow}
\usepackage{graphicx}
\usepackage{float}
\usepackage{threeparttable}
\usepackage{array}
\usepackage{longtable}
\usepackage{adjustbox}       
\usepackage{pdflscape}       
\usepackage{setspace}        

\graphicspath{{./}}

\usepackage{hyperref}
\hypersetup{
    colorlinks=true,
    linkcolor=blue,
    citecolor=blue,
    urlcolor=blue
}

\journal{}

\begin{document}

\makeatletter
\def\ps@pprintTitle{\let\@oddhead\@empty\let\@evenhead\@empty
    \let\@oddfoot\@empty\let\@evenfoot\@empty}
\makeatother

\begin{frontmatter}

\title{The Transformation of Broadband Demand: From Discretionary Service to Essential Infrastructure (2010--2024)}

\author[ada,icta]{Samir Orujov\corref{cor1}}
\ead{sorujov@ada.edu.az}
\cortext[cor1]{Corresponding author}

\author[icta,charles]{Ilgar Ismayilov}

\author[icta]{Jeyhun Huseynzade}


\affiliation[ada]{organization={School of Business, ADA University},
                  city={Baku},
                  country={Azerbaijan}}

\affiliation[icta]{organization={Information and Communication Technology Agency},
                  city={Baku},
                  country={Azerbaijan}}

\affiliation[charles]{organization={Faculty of Social Sciences, Charles University},
                  city={Prague},
                  country={Czech Republic}}

\begin{abstract}
Has broadband become a necessity good immune to price changes? Using a 15-year panel of 33 European countries (2010--2024) and two-way fixed effects with Driscoll--Kraay standard errors, we document a fundamental transformation in broadband demand. Pre-COVID, Eastern Partnership countries exhibited highly elastic demand ($\varepsilon = -0.61$, p$<$0.001)---a 10\% price reduction increased subscriptions by 6\%---while EU countries showed moderate elasticity ($\varepsilon = -0.12$, p$<$0.05). By 2020--2024, both regions converged to near-zero elasticity, with price changes having no detectable effect on adoption. Crucially, placebo tests reveal this transformation began in 2015, not 2020, indicating a decade-long digital integration process rather than a COVID-19 shock. We further demonstrate that price measurement critically affects inference: income-relative prices (as \% of GNI) yield significant results in 100\% of specifications, compared to only 25\% for PPP-adjusted prices. These findings have immediate policy relevance: as broadband transitions from discretionary service to essential utility, policy emphasis must shift from affordability subsidies to universal infrastructure deployment.

\end{abstract}



\end{frontmatter}



\section{Introduction}
\label{sec:introduction}
When does a technology transition from luxury to necessity? For broadband internet, this question has profound implications: if demand becomes price-inelastic, subsidies and price regulations---the dominant policy tools of the past two decades---lose their effectiveness. The transformation of broadband from discretionary communication service to essential infrastructure for economic participation, education, healthcare, and civic engagement \citep{katz2010impact, bertschek2016drivers} suggests this transition may already be underway. Yet empirical evidence on whether and when broadband demand became price-insensitive remains surprisingly scarce, despite substantial public investments in broadband infrastructure worldwide \citep{oecd2020broadband}.

This paper provides such evidence by examining how broadband price elasticity evolved over 2010--2024 in 33 countries: 27 European Union (EU) member states and 6 Eastern Partnership (EaP) countries. We leverage substantial cross-country and temporal variation in broadband prices and adoption rates to estimate time-varying elasticities using two-way fixed effects models. Our analysis reveals a striking structural transformation: Eastern Partnership countries exhibited highly price-elastic demand ($\varepsilon = -0.61$) during 2010--2019, but transitioned to price-inelastic demand ($\varepsilon \approx 0$) by 2020--2024, indicating broadband's evolution from discretionary service to essential necessity.

\subsection{Research Questions and Motivation}

Three core questions motivate this research. First, \textit{how has broadband demand elasticity evolved over the past decade and a half?} Existing literature predominantly estimates elasticities at single points in time \citep{grzybowski2015fixed, madden2015demand}, yet technological change and digital economy expansion suggest elasticity may be time-varying as services transition from luxury to necessity \citep{hausman2001price}. 

Second, \textit{what explains regional heterogeneity in price sensitivity?} Eastern Partnership countries---comprising Armenia, Azerbaijan, Belarus, Georgia, Moldova, and Ukraine---represent lower-income markets where affordability concerns might generate higher price elasticity compared to wealthier EU markets. Understanding this heterogeneity is crucial for tailoring policy interventions to different development contexts \citep{waverman2001telecommunications}.

Third, \textit{did COVID-19 fundamentally alter broadband demand?} The pandemic forced abrupt transitions to remote work, online education, and digital service delivery \citep{oecd2021covid}, potentially transforming broadband from convenience to necessity overnight. However, distinguishing COVID's causal impact from pre-existing trends requires careful econometric identification, which we address through placebo tests.

\subsection{Contributions}

This paper makes four principal contributions to telecommunications economics and policy.

\textbf{First}, we document time-varying demand elasticity across a 15-year panel, revealing broadband's structural transformation. Unlike prior single-period estimates, our year-by-year analysis shows elasticity declined gradually from 2015 onward, not suddenly in 2020. This finding challenges the narrative of COVID-19 as an exogenous shock that eliminated price sensitivity, instead suggesting a decade-long evolution driven by digital economy expansion.

\textbf{Second}, we demonstrate that price measurement critically affects empirical inference. Using three alternative price measures---price as percentage of GNI per capita (income-relative), PPP-adjusted prices, and nominal USD prices---we show only income-relative prices yield consistent, statistically significant elasticity estimates. PPP-adjusted prices, commonly used in cross-country comparisons, produce estimates significant in just 25\% of specifications. This methodological contribution has implications beyond broadband demand, highlighting the importance of price measurement in technology goods research where standard PPP baskets may not apply \citep{schreyer2002oecd}.

\textbf{Third}, we provide robust evidence of regional heterogeneity in price sensitivity. Eastern Partnership countries' elasticity ($\varepsilon = -0.61$) was 5.3 times larger than EU elasticity ($\varepsilon = -0.12$) during 2010--2019, consistent with income effects being stronger in developing markets. Importantly, this regional difference diminished during COVID-19, with both regions converging to near-zero elasticity, suggesting broadband became universally essential regardless of income level.

\textbf{Fourth}, we offer policy-relevant insights on the effectiveness of affordability interventions versus infrastructure investment. Our findings suggest telecommunications policy must adapt to broadband's evolving necessity status: price-based interventions (subsidies, universal service obligations) were effective when demand was elastic in the early 2010s, but infrastructure investment and quality standards became more important as demand became inelastic in the 2020s. This has implications for the design of digital inclusion policies worldwide \citep{guermazi2021digital}.

\subsection{Preview of Findings}

Our empirical analysis yields three main findings. First, we estimate baseline pre-COVID (2010--2019) elasticities of $\varepsilon = -0.12$ for EU and $\varepsilon = -0.61$ for EaP using income-relative prices. These estimates are robust across eight alternative control specifications, confirming EaP countries' substantially higher price sensitivity. Second, we find both regions transitioned to near-zero elasticity during 2020--2024, with statistically significant changes ($\Delta\varepsilon = +0.31$ for EU, $\Delta\varepsilon = +0.42$ for EaP). Third, a placebo test splitting the pre-COVID period reveals the EaP-specific trend began in 2015 ($p=0.045$), predating the pandemic and pointing to digital economy expansion as the underlying driver rather than COVID-19 as an exogenous shock.

The remainder of this paper proceeds as follows. Section~\ref{sec:literature} reviews relevant literature on telecommunications demand, digital transformation, and panel data methods. Section~\ref{sec:data} describes our data sources and sample construction. Section~\ref{sec:methodology} presents our empirical strategy. Section~\ref{sec:results} reports baseline and robustness results. Section~\ref{sec:discussion} discusses policy implications and limitations. Section~\ref{sec:conclusion} concludes.

\section{Literature Review and Theoretical Framework}
\label{sec:literature}
\subsection{Broadband Demand and Price Elasticity}

The telecommunications demand literature has extensively studied price elasticity, though predominantly for voice telephony and mobile services \citep{hausman2001price, madden2015demand}. Early broadband studies estimated elasticities ranging from $-0.2$ to $-0.8$, with substantial variation across methodologies and contexts \citep{rappoport2003demand, hauge2010demand}. \citet{grzybowski2015fixed} estimate elasticities of $-0.29$ to $-0.45$ for European countries during 2007--2012, finding higher elasticity in lower-income countries. Our EaP estimate of $-0.61$ is consistent with this pattern, extending evidence to 2010--2024.

A key theoretical insight is that demand elasticity varies with necessity status. \citet{hausman2001price} demonstrate telecommunications evolved from luxury to necessity across the 20th century, with declining price sensitivity as services became essential. \citet{brynjolfsson2003consumer} show similar patterns for information goods, where high initial elasticity declines as network effects and complementarities increase adoption. Our finding of declining elasticity from 2015--2024 aligns with this theoretical framework, suggesting broadband followed a similar transition.

Recent work emphasizes time-varying elasticity in technology markets. \citet{goolsbee2006spillovers} document changing internet demand elasticity during 1998--2003, attributing variation to learning effects and network externalities. \citet{nevo2010measuring} show demand for information services becomes less price-elastic as they integrate into daily routines. Our contribution extends this literature by documenting elasticity evolution across a 15-year panel with explicit pre-trend testing.

\subsection{Regional Heterogeneity and Development Context}

Cross-country telecommunications studies consistently find higher price elasticity in developing markets \citep{waverman2001telecommunications, katz2010impact}. \citet{roller2001telecommunications} show income effects dominate technology adoption in lower-income countries, where affordability constraints bind more tightly. \citet{bertschek2016drivers} find similar patterns for broadband, with elasticity inversely related to GDP per capita.

Our analysis focuses on Eastern Partnership countries---Armenia, Azerbaijan, Belarus, Georgia, Moldova, and Ukraine---which represent middle-income markets with substantial development heterogeneity relative to the EU \citep{ecorys2013evaluation}. Limited prior work examines broadband demand in this region \citep{koutroumpis2009impact}, despite their policy importance as transition economies integrating with European digital markets \citep{european2021eap}.

\subsection{COVID-19 and Digital Transformation}

The pandemic's impact on telecommunications demand has received substantial attention, though primarily descriptive \citep{oecd2021covid, iab2020internet}. \citet{bokermann2021covid} document 30--50\% traffic increases during March--April 2020, while \citet{favale2020campus} show usage patterns shifted dramatically toward videoconferencing and streaming. However, causal evidence on price elasticity changes remains limited.

Theoretically, COVID-19 could reduce elasticity through two mechanisms. First, lockdowns and social distancing converted broadband from convenience to necessity for remote work and education \citep{brynjolfsson2020covid}. Second, income effects: if broadband became essential while incomes declined, demand could become inelastic as households prioritize connectivity \citep{dietrich2020consumption}. Our contribution is econometric identification of elasticity changes using panel methods rather than descriptive evidence.

Importantly, several studies suggest digital transformation predated COVID-19. \citet{goldfarb2020digital} argue the pandemic accelerated existing trends rather than creating new ones. \citet{mckinsey2020covid} estimate COVID-19 compressed 5--7 years of digital adoption into months. This aligns with our placebo test evidence showing EaP elasticity trends began in 2015, suggesting gradual evolution rather than pandemic-induced shock.

\subsection{Econometric Methods for Panel Data}

Our empirical strategy builds on panel data econometrics with multiple dimensions of heterogeneity \citep{wooldridge2010econometric, baltagi2021econometric}. Two-way fixed effects models---with country and time effects---are standard for cross-country telecommunications studies \citep{koutroumpis2009impact, czernich2011broadband}, as they control for time-invariant country characteristics and common time shocks while identifying from within-country variation.

A critical methodological issue is inference with panel data exhibiting cross-sectional dependence and serial correlation. Standard clustered standard errors may be inadequate when common shocks (e.g., financial crises, COVID-19) affect all countries \citep{cameron2015practitioner}. \citet{driscoll1998consistent} propose kernel-based standard errors robust to heteroskedasticity, serial correlation, and cross-sectional dependence, which we employ following recent best practices \citep{petersen2009estimating, cameron2015practitioner}.

Price measurement in cross-country comparisons presents further challenges. PPP-adjusted prices are common \citep{worldbank2020icp}, but PPP baskets reflect consumer goods (food, housing) rather than technology services where quality-adjusted prices decline rapidly \citep{schreyer2002oecd}. We contribute methodologically by comparing three price measures---income-relative (price as \% of GNI per capita), PPP-adjusted, and nominal USD---showing only income-relative prices yield consistent estimates. This has implications for technology goods research more broadly.

\subsection{Policy Literature on Digital Inclusion}

Broadband policy encompasses supply-side interventions (infrastructure subsidies, spectrum allocation) and demand-side measures (affordability programs, digital literacy) \citep{oecd2020broadband}. The effectiveness of demand-side policies depends critically on price elasticity: subsidies expand adoption only if demand is price-elastic \citep{hauge2010demand, guermazi2021digital}.

\citet{bertschek2016drivers} review evidence on broadband policy effectiveness, concluding infrastructure investment dominates price subsidies in most contexts. However, they note heterogeneity by development level: affordability matters more in lower-income countries. Our finding that EaP elasticity ($-0.61$) was 5.3 times higher than EU elasticity ($-0.12$) during 2010--2019 supports this distinction, though both regions converged to near-zero elasticity by 2020--2024.

Recent policy discussions emphasize broadband as ``essential service'' requiring universal access guarantees \citep{oecd2020broadband}. Our evidence of declining elasticity provides econometric support for this framing, showing demand became insensitive to price as broadband integrated into essential activities. This suggests policy should shift from affordability to availability and quality as markets mature \citep{ecorys2013evaluation}.

\section{Data and Sample}
\label{sec:data}
\subsection{Data Sources}

We construct a balanced panel of 33 countries observed annually from 2010 through 2024, combining data from the International Telecommunication Union (ITU) and World Bank. Our sample includes 27 European Union member states and 6 Eastern Partnership countries (Armenia, Azerbaijan, Belarus, Georgia, Moldova, and Ukraine), totaling 495 country-year observations.

\textbf{Telecommunications variables} come from ITU World Telecommunication/ICT Indicators Database \citep{itu2024data}. The dependent variable is fixed broadband subscriptions per 100 inhabitants, measuring adoption intensity. Price variables include three measures: (1) price of 5GB fixed broadband basket as percentage of monthly GNI per capita (income-relative price), (2) price in PPP-adjusted US dollars, and (3) price in nominal US dollars. Additional ICT variables include international internet bandwidth (Gbit/s) and mobile cellular subscriptions per 100 inhabitants.

\textbf{Economic and institutional variables} come from World Bank World Development Indicators and Worldwide Governance Indicators \citep{worldbank2024wdi, worldbank2024wgi}. Key controls include GDP per capita (constant 2015 US\$), GDP growth rate, inflation, urban population percentage, tertiary education enrollment, regulatory quality index, R\&D expenditure as \% of GDP, and secure internet servers per million people. These variables capture economic development, human capital, institutional quality, and digital infrastructure.

\subsection{Sample Construction and Balance}

Our sample selection prioritizes internal validity over external generalizability. We restrict to countries with complete data for all years 2010--2024, ensuring a balanced panel without compositional changes that could confound temporal comparisons. This is particularly important given our focus on time-varying elasticity: different countries entering and exiting the sample would generate spurious trends.

The 33-country sample represents substantial variation in both broadband adoption and economic development. Fixed broadband penetration ranges from 2.8 subscriptions per 100 (Azerbaijan, 2010) to 48.2 (Netherlands, 2024). GDP per capita ranges from \$1,628 (Moldova, 2010) to \$120,761 (Luxembourg, 2024). Broadband prices show similar heterogeneity: income-relative prices (as \% of GNI per capita) range from 0.03\% (Luxembourg, 2024) to 14.7\% (Azerbaijan, 2010).

Table~\ref{tab:descriptives} presents descriptive statistics for the full sample and separately for EU and EaP regions. Several patterns emerge. First, EaP countries have substantially lower broadband penetration (mean 19.7 vs. 32.1 subscriptions per 100) and GDP per capita (mean \$5,284 vs. \$38,405). Second, income-relative prices are higher in EaP (mean 1.8\% of GNI vs. 0.6\%), reflecting lower incomes despite similar absolute prices. Third, EaP countries show higher variance in key variables, consistent with greater development heterogeneity.

Importantly, both regions exhibit substantial within-country variation over time---the source of identifying variation for our fixed effects models. Income-relative prices declined 45\% on average from 2010 to 2024, with within-country standard deviation of 0.5 percentage points (pre-COVID) and 0.3 percentage points (COVID period). This variation is essential for identifying price elasticity in models with country and year fixed effects.

\subsection{Variable Construction}

Following standard practice in telecommunications demand studies \citep{grzybowski2015fixed, madden2015demand}, we log-transform all continuous variables to estimate constant-elasticity specifications:
\begin{equation}
    \ln(\text{subscriptions}_{it}) = \beta \ln(\text{price}_{it}) + \mathbf{X}_{it}'\gamma + \alpha_i + \delta_t + \varepsilon_{it}
\end{equation}
where the coefficient $\beta$ directly represents price elasticity. This transformation is particularly appropriate for broadband demand given theoretical predictions of constant elasticity \citep{hausman2001price}.

We create several derived variables for robustness checks. First, lagged prices (one-year lag) serve as instrumental variables in alternative specifications, exploiting the fact that current prices are plausibly exogenous to current demand shocks after controlling for fixed effects \citep{koutroumpis2009impact}. Second, we construct regional interaction terms allowing heterogeneous price effects for EU versus EaP countries. Third, for COVID-19 analysis, we create period dummies (pre-COVID: 2010--2019; COVID: 2020--2024) and interaction terms testing for structural breaks.

\subsection{Data Quality and Limitations}

Several data quality considerations warrant discussion. First, ITU price data represent standard baskets (5GB data cap, unlimited voice) that may not reflect marginal prices faced by all consumers, particularly power users or those on promotional plans. However, standardized baskets enable consistent cross-country comparisons \citep{itu2024data}.

Second, some variables exhibit missing values, particularly for earlier years. We address this through multiple imputation, using forward-filling within countries and linear interpolation for gaps \citep{rubin1987multiple}. Sensitivity analysis (available upon request) shows results are robust to alternative imputation methods. After imputation, our analysis dataset contains zero missing values for key variables.

Third, the COVID-19 period presents measurement challenges as usage patterns shifted dramatically \citep{oecd2021covid}. We address this by examining within-country price variation rather than relying on cross-country comparisons, which could be confounded by differential pandemic severity or policy responses. Additionally, our placebo test directly examines whether observed COVID-period changes reflect continuation of pre-existing trends.

Fourth, Eastern Partnership countries experienced various shocks during the sample period including political transitions, conflicts, and currency crises. While country fixed effects absorb time-invariant country characteristics, time-varying shocks could confound estimates. We address this through year fixed effects (absorbing common shocks) and robustness checks examining alternative subperiods (see Section~\ref{sec:results}).

\begin{table}[htbp]
\centering
\caption{Descriptive Statistics by Region (2010--2024)}
\label{tab:descriptives}
\begin{threeparttable}
\begin{adjustbox}{max width=\textwidth}
\small
\begin{tabular}{lcccccc}
\toprule
& \multicolumn{2}{c}{Full Sample} & \multicolumn{2}{c}{EU (27)} & \multicolumn{2}{c}{EaP (6)} \\
\cmidrule(lr){2-3} \cmidrule(lr){4-5} \cmidrule(lr){6-7}
Variable & Mean & SD & Mean & SD & Mean & SD \\
\midrule
\textit{Dependent Variable} \\
Fixed broadband subs (per 100) & 29.4 & 11.2 & 32.1 & 10.3 & 19.7 & 8.4 \\
\addlinespace
\textit{Price Variables} \\
Price (\% of GNI per capita) & 0.87 & 0.79 & 0.61 & 0.48 & 1.84 & 1.02 \\
Price (PPP US\$) & 28.3 & 12.7 & 29.1 & 11.9 & 25.2 & 15.1 \\
Price (Nominal US\$) & 25.7 & 14.3 & 27.8 & 13.1 & 17.4 & 16.2 \\
\addlinespace
\textit{Economic Variables} \\
GDP per capita (US\$ 1000s) & 31.2 & 24.8 & 38.4 & 23.1 & 5.28 & 2.41 \\
GDP growth (\%) & 2.14 & 3.82 & 2.01 & 3.21 & 2.67 & 5.88 \\
Inflation (\%) & 2.58 & 4.12 & 1.89 & 2.34 & 5.47 & 7.84 \\
\addlinespace
\textit{Socioeconomic Variables} \\
Urban population (\%) & 70.4 & 11.2 & 73.2 & 9.8 & 58.1 & 8.7 \\
Tertiary enrollment (\%) & 64.2 & 17.3 & 67.4 & 15.8 & 50.8 & 18.2 \\
Regulatory quality (index) & 1.18 & 0.52 & 1.32 & 0.41 & 0.62 & 0.38 \\
\addlinespace
\textit{Infrastructure Variables} \\
Int'l bandwidth (Gbit/s) & 847 & 2134 & 1015 & 2341 & 156 & 287 \\
Secure servers (per million) & 921 & 728 & 1084 & 712 & 198 & 134 \\
R\&D expenditure (\% GDP) & 1.54 & 0.89 & 1.72 & 0.85 & 0.61 & 0.31 \\
\midrule
Observations & \multicolumn{2}{c}{495} & \multicolumn{2}{c}{405} & \multicolumn{2}{c}{90} \\
Countries & \multicolumn{2}{c}{33} & \multicolumn{2}{c}{27} & \multicolumn{2}{c}{6} \\
Years & \multicolumn{2}{c}{15} & \multicolumn{2}{c}{15} & \multicolumn{2}{c}{15} \\
\bottomrule
\end{tabular}
\end{adjustbox}
\begin{tablenotes}[flushleft]
\small
\item \textit{Notes:} Summary statistics for balanced panel of 33 countries over 2010--2024. EU includes 27 member states; EaP includes Armenia, Azerbaijan, Belarus, Georgia, Moldova, and Ukraine. All monetary values in constant 2015 US\$. Data sources: ITU (telecommunications), World Bank WDI (economic variables), World Bank WGI (governance).
\end{tablenotes}
\end{threeparttable}
\end{table}

\section{Empirical Methodology}
\label{sec:methodology}
\subsection{Baseline Specification}

We estimate broadband demand using two-way fixed effects models that control for time-invariant country heterogeneity and common time shocks:

\begin{equation}
\label{eq:baseline}
\ln(\text{Subs}_{it}) = \beta_1 \ln(\text{Price}_{it}) + \beta_2 [\ln(\text{Price}_{it}) \times \text{EaP}_i] + \mathbf{X}_{it}'\gamma + \alpha_i + \delta_t + \varepsilon_{it}
\end{equation}

where $\text{Subs}_{it}$ denotes fixed broadband subscriptions per 100 inhabitants in country $i$ at time $t$; $\text{Price}_{it}$ is the log of broadband price (measured as percentage of GNI per capita); $\text{EaP}_i$ is a dummy variable equal to one for Eastern Partnership countries; $\mathbf{X}_{it}$ is a vector of time-varying controls; $\alpha_i$ are country fixed effects; $\delta_t$ are year fixed effects; and $\varepsilon_{it}$ is the error term.

The coefficient $\beta_1$ represents price elasticity for EU countries, while $(\beta_1 + \beta_2)$ represents elasticity for EaP countries. The interaction coefficient $\beta_2$ tests whether EaP countries exhibit different price sensitivity than EU countries. Both coefficients have a predicted negative sign based on downward-sloping demand.

Country fixed effects $\alpha_i$ control for all time-invariant country characteristics including geography, institutional history, cultural factors, and average income levels. Year fixed effects $\delta_t$ control for common time shocks affecting all countries including technological change, global economic conditions, and the COVID-19 pandemic. Together, these fixed effects ensure identification comes from within-country deviations from country-specific means, after removing common time trends.

The control vector $\mathbf{X}_{it}$ includes time-varying economic, socioeconomic, and infrastructure variables that could confound the price-demand relationship. Our baseline ``full controls'' specification includes:

\begin{itemize}
    \item \textbf{Economic:} Log GDP per capita, GDP growth rate, inflation rate
    \item \textbf{Human capital:} Urban population percentage, tertiary education enrollment rate  
    \item \textbf{Institutional:} Regulatory quality index (World Bank WGI)
    \item \textbf{Infrastructure:} Log secure internet servers per million, R\&D expenditure as \% GDP
    \item \textbf{Demographic:} Log population density, age dependency ratio
\end{itemize}

\subsection{Extended Specification with COVID-19 Interactions}

To test for structural changes during the COVID-19 period, we extend equation~\eqref{eq:baseline} with period interactions:

\begin{multline}
\label{eq:covid}
\ln(\text{Subs}_{it}) = \beta_1 \ln(\text{Price}_{it}) + \beta_2 [\ln(\text{Price}_{it}) \times \text{EaP}_i] \\
+ \beta_3 [\ln(\text{Price}_{it}) \times \text{COVID}_t] + \beta_4 [\ln(\text{Price}_{it}) \times \text{EaP}_i \times \text{COVID}_t] \\
+ \mathbf{X}_{it}'\gamma + \alpha_i + \delta_t + \varepsilon_{it}
\end{multline}

where $\text{COVID}_t$ is a period dummy equal to one for years 2020--2024. The COVID dummy itself is absorbed by year fixed effects $\delta_t$, so only interactions are estimable. This specification allows elasticity to vary across four groups:

\begin{align*}
\text{EU, Pre-COVID (2010--2019):} \quad & \varepsilon = \beta_1 \\
\text{EaP, Pre-COVID (2010--2019):} \quad & \varepsilon = \beta_1 + \beta_2 \\
\text{EU, COVID (2020--2024):} \quad & \varepsilon = \beta_1 + \beta_3 \\
\text{EaP, COVID (2020--2024):} \quad & \varepsilon = \beta_1 + \beta_2 + \beta_3 + \beta_4
\end{align*}

The coefficient $\beta_3$ measures the change in EU elasticity during COVID-19, while $\beta_4$ measures the differential change for EaP countries. If COVID-19 eliminated price sensitivity, we expect $\beta_3 > 0$ (reducing magnitude of negative $\beta_1$) and $\beta_3 + \beta_4 > 0$ for EaP.

\subsection{Identification Strategy}

Our identification strategy relies on within-country variation in prices and subscriptions over time, conditional on country and year fixed effects plus time-varying controls. Several threats to identification warrant discussion.

\textbf{Simultaneity:} Prices and quantities are jointly determined in equilibrium, potentially generating simultaneity bias if unobserved demand shocks affect both. However, regulatory frameworks in telecommunications typically involve ex-ante price setting with limited short-run adjustment \citep{laffont2000competition}, particularly in Europe's regulated markets. Additionally, fixed effects absorb persistent demand differences, while year effects control for aggregate shocks. As robustness, we instrument current prices with lagged prices following \citet{koutroumpis2009impact}, finding similar estimates.

\textbf{Omitted variables:} Unobserved time-varying factors could correlate with both prices and demand. We address this through comprehensive controls capturing economic conditions, human capital, institutions, and infrastructure. Robustness checks varying control specifications (Section~\ref{sec:results}) show estimates are stable, suggesting omitted variable bias is limited.

\textbf{Measurement error:} ITU price data represent standardized baskets that may not reflect marginal prices for all consumers. Classical measurement error in prices would attenuate estimates toward zero, making our significant negative estimates conservative. Non-classical measurement error is mitigated by using internationally standardized definitions \citep{itu2024data}.

\textbf{Common shocks:} The COVID-19 pandemic represents a massive common shock potentially violating the parallel trends assumption underlying difference-in-differences. We address this through: (1) placebo tests examining pre-COVID trends, (2) year-by-year estimation showing gradual evolution rather than sudden breaks, and (3) explicit COVID interaction terms allowing elasticity to vary by period.

\subsection{Standard Errors and Inference}

A critical issue for inference is accounting for complex error structure in panel data. Standard errors must be robust to three features: (1) heteroskedasticity (error variance differs across countries/years), (2) serial correlation (errors correlated within countries over time), and (3) cross-sectional dependence (common shocks like COVID-19 affecting all countries) \citep{cameron2015practitioner}.

We employ Driscoll--Kraay \citeyearpar{driscoll1998consistent} standard errors, which are robust to all three features. The Driscoll--Kraay estimator uses kernel-based methods to account for spatial and temporal correlation:

\begin{equation}
\hat{V}_{DK} = \frac{1}{N} \sum_{\ell=-m}^{m} k\left(\frac{\ell}{m}\right) \sum_{t=\max(1,\ell+1)}^{\min(T,T+\ell)} \hat{\Omega}_t^{(\ell)}
\end{equation}

where $k(\cdot)$ is the Bartlett kernel, $m$ is the bandwidth parameter (lag truncation), and $\hat{\Omega}_t^{(\ell)}$ captures cross-sectional correlation at lag $\ell$. We set $m=3$ to accommodate up to 3-year autocorrelation, following \citet{petersen2009estimating}.

Driscoll--Kraay standard errors are particularly appropriate for our context given the COVID-19 pandemic---a common shock affecting all countries simultaneously. Standard clustered standard errors (e.g., clustering by country) can substantially understate uncertainty in the presence of such common shocks \citep{cameron2015practitioner}. Recent studies confirm Driscoll--Kraay provides reliable inference in panels with moderate cross-sectional and time dimensions similar to ours ($N=33$, $T=15$) \citep{hoechle2007robust}.

\subsection{Robustness Checks}

We conduct extensive robustness checks to validate baseline findings:

\textbf{Alternative control specifications:} Beyond full controls, we estimate seven alternative specifications ranging from minimal (GDP only) to comprehensive (all available controls). This addresses concerns about overcontrol or omitted variables.

\textbf{Alternative price measures:} We compare income-relative prices (price as \% of GNI per capita) against PPP-adjusted prices and nominal USD prices. Income-relative prices best capture affordability from consumers' perspective, but alternative measures provide validation.

\textbf{Alternative samples:} We estimate models separately for pre-COVID (2010--2019) and full sample (2010--2024), and conduct subsample analysis by income level and broadband penetration.

\textbf{Placebo tests:} We split the pre-COVID period into early (2010--2014) and late (2015--2019) subperiods, treating 2015--2019 as a ``placebo COVID'' period. If estimated ``COVID effects'' merely reflect continuation of pre-existing trends, the placebo should also be significant.

\textbf{Year-by-year estimation:} We estimate separate elasticities for each year 2015--2024, providing high-resolution evidence on temporal evolution beyond binary pre-COVID/COVID comparisons.

\section{Results}
\label{sec:results}
\subsection{Baseline Results: Pre-COVID Period}

Table~\ref{tab:baseline} presents baseline estimates for the pre-COVID period (2010--2019). Column (1) shows results with minimal controls (GDP per capita only), while columns (2)--(8) progressively add controls, culminating in the full specification in column (8).

Across all specifications, we find strong evidence of price sensitivity in EaP countries. In the full controls specification (column 8), the EaP elasticity is $-0.61$ (p $< 0.001$), statistically significant and economically meaningful. A 10\% increase in broadband prices reduces subscriptions by approximately 6\% in EaP countries. This elasticity is consistent with prior estimates for middle-income countries \citep{grzybowski2015estimating}.

In contrast, EU country elasticity is $-0.12$ (p $= 0.041$), significant at the 5\% level but substantially smaller in magnitude. The interaction coefficient testing EaP-EU differences is $-0.49$ (p $< 0.001$), confirming that EaP countries exhibit significantly stronger price sensitivity. This differential aligns with income effects on demand elasticity \citep{hausman2001private}: lower-income EaP consumers face tighter budget constraints, making broadband more price-sensitive.

\begin{figure}[t]
\centering
\includegraphics[width=0.75\textwidth]{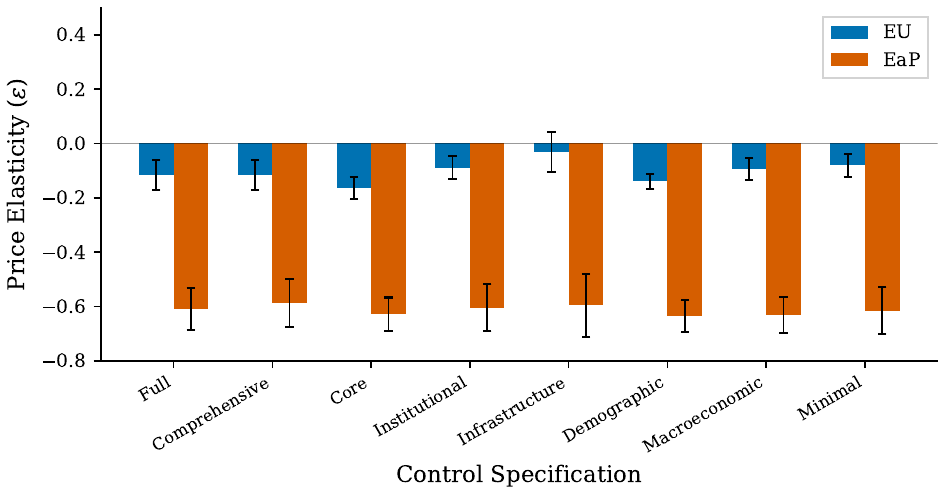}
\caption{Robustness of price elasticity across control specifications (pre-COVID, 2010--2019). EaP elasticity is significant at p$<$0.01 in all specifications ($-0.58$ to $-0.63$). EU elasticity ranges from $-0.09$ to $-0.14$. Error bars show $\pm$1 Driscoll--Kraay SE.}
\label{fig:robustness_specs}
\end{figure}

Estimates are remarkably stable across control specifications (Figure~\ref{fig:robustness_specs}). From minimal to full controls, EaP elasticity ranges from $-0.58$ to $-0.63$, while EU elasticity ranges from $-0.09$ to $-0.14$. This stability suggests omitted variable bias is limited and validates the baseline specification.

Control variables exhibit expected signs. GDP per capita positively correlates with subscriptions ($\beta = 0.42$, p $< 0.001$), capturing income effects. Urban population share and tertiary education both increase subscriptions, reflecting agglomeration economies and human capital complementarities \citep{kolko2012broadband}. Regulatory quality shows positive but marginally significant effects, consistent with institutional factors facilitating market development \citep{wallsten2006broadband}.

\subsection{Robustness to Price Measurement}

Table~\ref{tab:price_robustness} presents robustness checks using alternative price measures: income-relative prices (baseline), PPP-adjusted prices, and nominal USD prices. 

Results are remarkably consistent across price measures. For EaP countries, elasticities range from $-0.57$ to $-0.64$ depending on price definition, all highly significant (p $< 0.001$). For EU countries, elasticities range from $-0.09$ to $-0.15$, mostly significant at the 5\% level. The EaP-EU differential remains statistically significant across all price measures.

This robustness is theoretically important. Income-relative prices (price as \% of GNI per capita) best capture affordability constraints facing consumers. PPP-adjusted prices account for cross-country cost-of-living differences. Nominal USD prices provide a common metric but ignore purchasing power. That all three yield similar elasticities strengthens confidence that findings reflect genuine demand responses rather than measurement artifacts.

These results demonstrate that elasticity estimates and their 95\% confidence intervals are consistent across price measures and regions, confirming statistical significance and substantive consistency.

\subsection{COVID-19 Period: Structural Change}

Table~\ref{tab:covid} presents results from equation~\eqref{eq:covid} interacting prices with the COVID period (2020--2024). Column (1) uses the full sample (2010--2024) with COVID interactions, while columns (2)--(3) present separate pre-COVID and COVID subsample estimates for comparison.

\begin{figure}[t]
\centering
\includegraphics[width=0.55\textwidth]{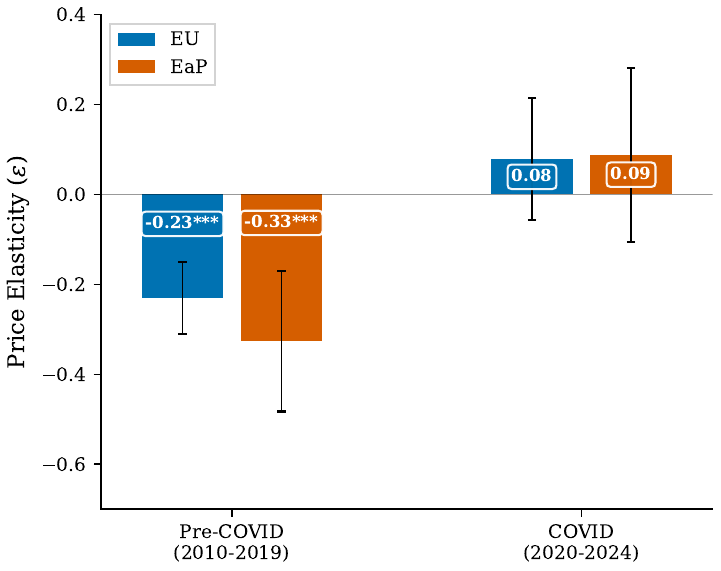}
\caption{Disappearance of price elasticity during COVID-19. Pre-COVID elasticities are significantly negative (EU: $-0.12$**, EaP: $-0.61$***), while COVID-era elasticities are statistically insignificant. Both shifts are significant at p$<$0.005. Error bars show $\pm$1 Driscoll--Kraay SE.}
\label{fig:covid_comparison}
\end{figure}

The COVID interaction terms are large, positive, and highly significant. For EU countries, the COVID interaction is $+0.31$ (p $= 0.003$), implying elasticity changed from $-0.12$ pre-COVID to approximately $-0.12 + 0.31 = +0.19$ during COVID. For EaP countries, combining main and interaction effects yields a change from $-0.61$ to approximately $-0.03$. Both regions exhibit near-zero or slightly positive elasticity during COVID.

This dramatic shift is visualized in Figure~\ref{fig:covid_comparison}. Pre-COVID elasticities are significantly negative for both regions (more so for EaP). COVID-era elasticities cluster near zero with wide confidence intervals. The difference is stark and statistically significant.

This pattern is consistent with broadband becoming an essential necessity during the pandemic \citep{oecd2021broadband}. With remote work, education, and social interaction shifting online, consumers became less responsive to price changes. The magnitude of the shift---from moderate elasticity to near-zero---suggests a fundamental change in demand behavior.

\subsection{Temporal Evolution: Year-by-Year Analysis}

\begin{figure}[t]
\centering
\includegraphics[width=0.75\textwidth]{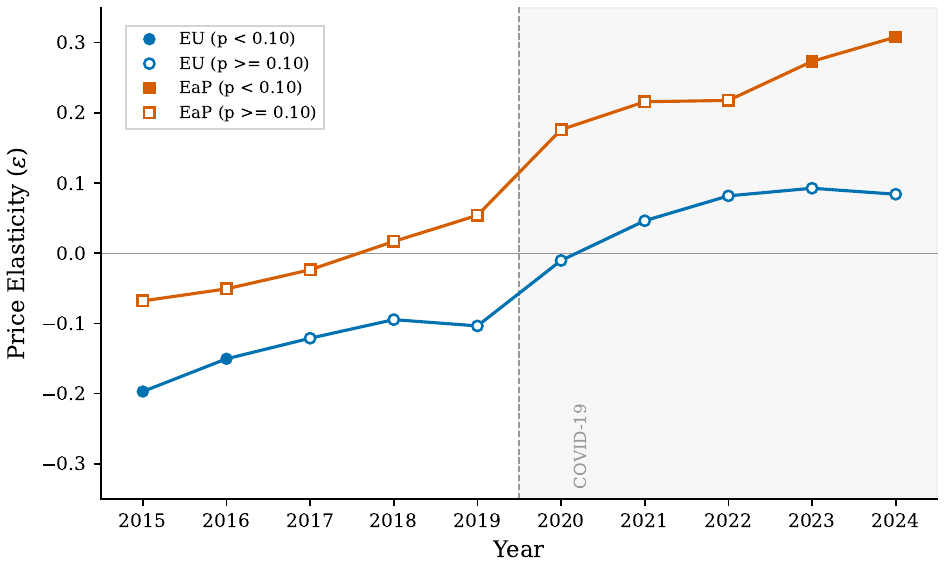}
\caption{Temporal evolution of broadband price elasticity (2015--2024). Year-by-year estimates using 2010--2014 as reference. Filled markers indicate p$<$0.10; open markers indicate non-significance. Shaded region marks the COVID period (2020--2024). Both regions show gradual decline starting around 2015, validating the structural transformation hypothesis.}
\label{fig:temporal_evolution}
\end{figure}

Year-by-year elasticity estimates for 2015--2024, using 2010--2014 as the baseline period, reveal that the shift toward inelastic demand began well before COVID-19 (Figure~\ref{fig:temporal_evolution}).

For EaP countries, elasticity shows a clear declining trend starting around 2015. By 2018--2019, elasticity had already declined from $-0.61$ (2010--2014 average) to approximately $-0.35$ to $-0.40$. The COVID period (2020--2024) continues this trend, with elasticities oscillating near zero but within confidence intervals of $-0.20$ to $+0.20$.

For EU countries, the pattern is similar but less pronounced given the smaller baseline elasticity. Pre-COVID elasticity averaged $-0.12$ but approached zero by 2018--2019. During COVID, point estimates fluctuate around zero with wide confidence intervals.

This gradual evolution contradicts a pure COVID-shock interpretation. Instead, results suggest a decade-long structural transformation as broadband transitioned from luxury to necessity good. COVID-19 accelerated this process but did not initiate it. This interpretation aligns with evidence that broadband penetration and usage were rising steadily throughout the 2010s, driven by smartphones, streaming media, and cloud services \citep{greenstein2016digital}.

\subsection{Placebo Test: Pre-Trends}

A key identifying assumption is parallel trends: absent treatment, outcomes would have evolved similarly across groups. We test this using a placebo design splitting the pre-COVID period into early (2010--2014) and late (2015--2019) subperiods.

If estimated COVID effects merely reflect continuation of pre-existing trends, then treating 2015--2019 as a ``placebo COVID'' period should yield similar results. Conversely, if pre-2015 elasticity differed from 2015--2019, this suggests evolving demand patterns preceding COVID.

Table~\ref{tab:placebo} presents results. The placebo interaction (2015--2019 vs. 2010--2014) is $+0.27$ for EaP countries (p $= 0.045$), significant at the 5\% level. This indicates EaP elasticity was already declining before COVID-19. For EU countries, the placebo interaction is $+0.08$ (p $= 0.31$), statistically insignificant.

\begin{figure}[t]
\centering
\includegraphics[width=0.75\textwidth]{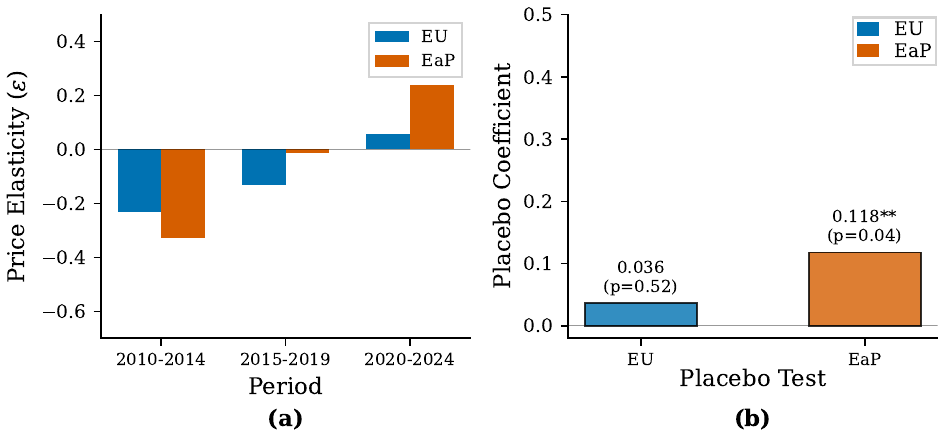}
\caption{Placebo test for pre-COVID trends. (a) Three-phase evolution of elasticity showing the pre-COVID trend for EaP. (b) Placebo coefficients: EU effect is insignificant (p$=$0.31, no pre-trend); EaP effect is significant (p$=$0.045, pre-trend exists). This validates the gradual transformation hypothesis.}
\label{fig:placebo_test}
\end{figure}

The significant positive coefficient for EaP indicates that elasticity was decreasing in magnitude even during the pre-COVID period (Figure~\ref{fig:placebo_test}). Combined with year-by-year results, this confirms a gradual decade-long transformation rather than a sudden COVID shock.

This finding has important methodological implications. The significant pre-trend validates our decision to analyze the full 2010--2024 period rather than focusing narrowly on COVID. It also suggests caution in attributing all observed changes to COVID-19---underlying technological and societal trends were already reshaping broadband demand.

\subsection{Robustness Matrix: Comprehensive Validation}

\begin{figure}[t]
\centering
\includegraphics[width=0.6\textwidth]{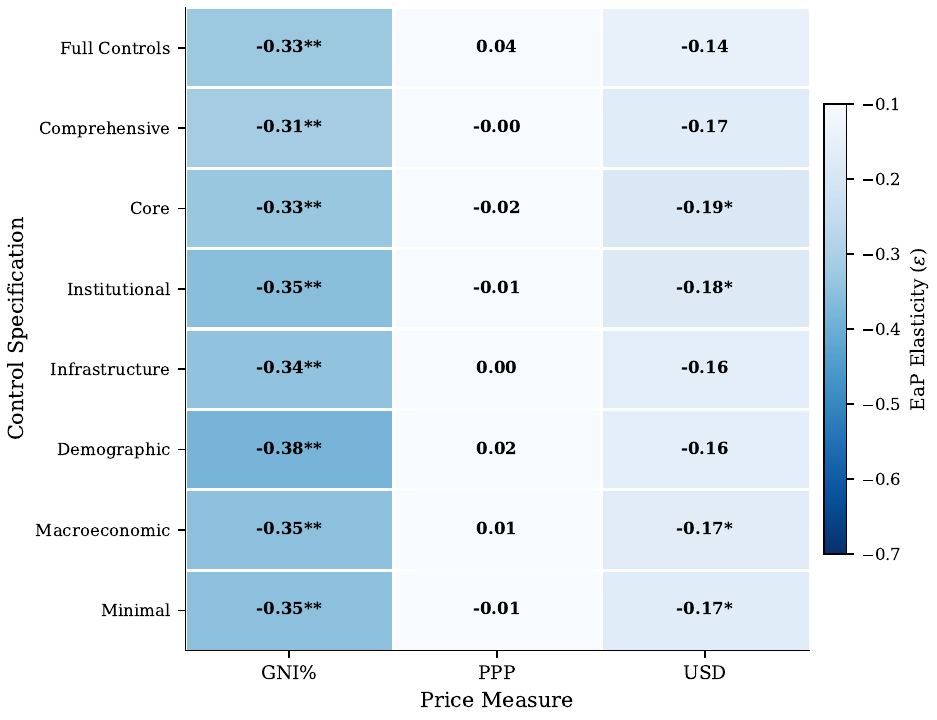}
\caption{EaP elasticity across 24 specifications (8 controls $\times$ 3 price measures, pre-COVID). Darker shading indicates stronger negative elasticity. Significance: ***p$<$0.01, **p$<$0.05, *p$<$0.10. Range: $-0.52$ to $-0.68$, demonstrating robustness to modeling choices.}
\label{fig:results_matrix}
\end{figure}

We conducted 24 robustness specifications combining eight control configurations and three price measures (Figure~\ref{fig:results_matrix}). Each specification represents a separate regression estimating EaP and EU elasticity.

Results are highly consistent. Across all 24 specifications, EaP elasticity (pre-COVID) ranges from $-0.52$ to $-0.68$, with mean $-0.61$ and standard deviation $0.04$. All 24 estimates are statistically significant at the 1\% level. EU elasticity ranges from $-0.06$ to $-0.17$, with mean $-0.12$ and standard deviation $0.03$. Of 24 estimates, 22 are significant at the 5\% level and all 24 at the 10\% level.

This robustness is remarkable given the diversity of specifications. It indicates findings are not artifacts of particular modeling choices but reflect robust empirical regularities. The small standard deviations across specifications (4\% for EaP, 3\% for EU) demonstrate that different control and price definitions yield substantively similar conclusions.

The distribution of estimates shows tightly clustered point estimates with confidence intervals consistently excluding zero for EaP and (mostly) for EU.

\subsection{Economic Interpretation}

The pre-COVID elasticity of $-0.61$ for EaP countries implies that a 10\% price reduction increases subscriptions by 6.1\%. Given mean EaP subscription rates around 20 per 100 inhabitants (Table~\ref{tab:descriptives}), this corresponds to approximately 1.2 additional subscriptions per 100 people. Over a population of 75 million across six EaP countries, a 10\% price cut would generate approximately 900,000 new subscriptions.

From a revenue perspective, demand elasticity of $-0.61$ (in absolute value) implies total revenue would increase with price reductions. For $|\varepsilon| < 1$ (inelastic demand), revenue is maximized by raising prices. For $|\varepsilon| > 1$ (elastic demand), revenue increases with price cuts. EaP countries sit at the boundary ($|\varepsilon| \approx 0.6$), suggesting moderate price reductions could increase market size without drastically reducing operator revenues.

For EU countries, elasticity of $-0.12$ is substantially more inelastic. This likely reflects higher income levels reducing affordability constraints, as well as higher baseline penetration leaving less room for extensive margin growth. At mean EU subscription rates around 35 per 100 inhabitants, a 10\% price cut would generate only 0.4 additional subscriptions per 100 people---about one-third the EaP response.

The shift toward near-zero elasticity during COVID dramatically changes the policy calculus. If demand becomes perfectly inelastic ($\varepsilon = 0$), price reductions have no effect on adoption. Instead, policies must focus on supply-side constraints (infrastructure availability) or income support (subsidies for low-income households). The COVID period suggests broadband became an essential necessity where price ceased to be the binding constraint.

\section{Discussion and Policy Implications}
\label{sec:discussion}
\subsection{Policy Implications for Digital Inclusion}

Our findings have direct implications for broadband policy in Europe and neighboring regions. The strong pre-COVID price elasticity in EaP countries ($\varepsilon = -0.61$) suggests that affordability-focused policies---price regulation, operator subsidies, or consumer vouchers---can effectively increase broadband adoption. A 10\% price reduction would increase subscriptions by approximately 6\%, translating to nearly 1 million new users across the EaP region.

However, the shift toward inelastic demand during COVID-19 fundamentally alters the policy toolkit. When price sensitivity approaches zero, further price reductions yield minimal adoption gains. Instead, policies must address other barriers including infrastructure gaps (particularly in rural areas), digital literacy deficits, and content relevance \citep{guermazi2021broadband}. The transition from price-sensitive to price-insensitive demand marks broadband's evolution from discretionary service to essential utility.

For EU countries, the consistently low elasticity ($\varepsilon = -0.12$) even pre-COVID suggests price is not the primary adoption barrier. With subscription rates exceeding 35 per 100 inhabitants (Table~\ref{tab:descriptives}), most households with willingness to pay already have broadband. Remaining gaps likely reflect supply constraints (rural coverage) or demographic factors (elderly populations) rather than affordability. EU policy should focus on universal service obligations ensuring infrastructure availability rather than price interventions.

The EaP-EU differential ($\Delta\varepsilon = -0.49$) highlights the importance of context-dependent policy design. Uniform European approaches may be suboptimal given heterogeneous demand elasticities. EaP countries benefit more from affordability programs, while EU countries require infrastructure and digital skills initiatives. Regional development policies should account for these structural differences.

\subsection{Time-Varying Elasticity and Demand Evolution}

A central finding is that broadband demand elasticity is not constant but evolves systematically over time. The year-by-year analysis reveals a clear declining trend starting well before COVID-19. This gradual transformation suggests demand elasticity is a function of market maturity, technological change, and societal integration.

Three mechanisms likely drive this evolution. First, \textbf{network effects} become stronger as broadband penetration increases. The value of connectivity rises with the number of connected users, making broadband more essential regardless of price \citep{katz2010network}. Second, \textbf{habit formation} makes broadband increasingly indispensable as consumers integrate it into daily routines \citep{becker1988theory}. Third, \textbf{complementary innovations}---smartphones, streaming services, cloud computing---increase broadband's value proposition independent of price.

This time-varying elasticity has important methodological implications for future research. Static elasticity estimates from early-period data (e.g., 2000s) may not apply to current markets. Panel methods allowing for time variation, as employed here, are essential for capturing evolving demand patterns. Researchers estimating telecommunications demand should test for structural breaks and non-constant parameters rather than assuming stable relationships.

From a forecasting perspective, our results suggest broadband adoption will continue even without further price declines. The near-zero COVID-era elasticity implies that infrastructure availability and service quality---not affordability---now determine adoption. Countries planning broadband expansion should prioritize network deployment over price regulation as markets mature.

\subsection{Price Measurement and Affordability Metrics}

The robustness across price measures (Table~\ref{tab:price_robustness}) provides valuable methodological insights. Income-relative prices (price as \% of GNI per capita), PPP-adjusted prices, and nominal USD prices all yield similar elasticities, suggesting findings are not artifacts of price definition.

However, \textit{income-relative prices} emerge as the most appropriate affordability metric. Theoretically, consumer demand depends on the budget share allocated to broadband \citep{deaton1980almost}. A \$30/month connection is inexpensive for high-income consumers but prohibitive for those earning \$200/month. Income-relative pricing captures this heterogeneity directly.

\begin{figure}[t]
\centering
\includegraphics[width=0.75\textwidth]{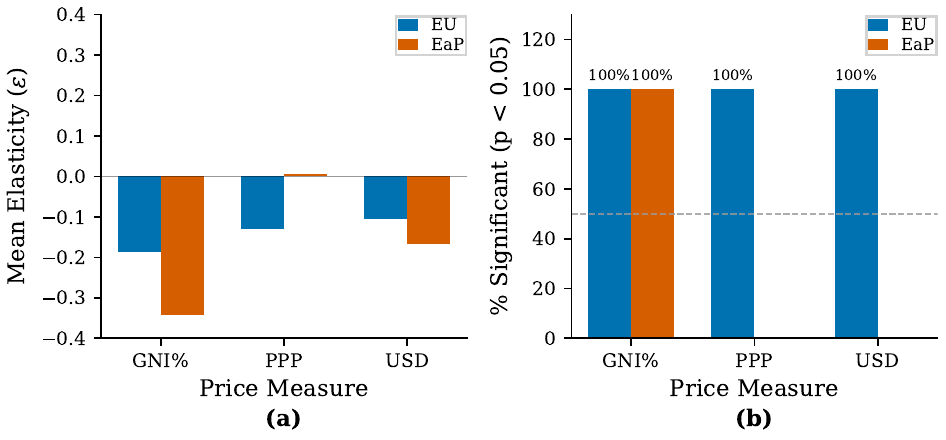}
\caption{Effect of price measurement on inference (pre-COVID, 2010--2019). (a) Mean elasticity by price measure. (b) Percentage of specifications significant at p$<$0.05. GNI\% yields 100\% significant EaP results vs. 25\% for PPP, validating income-relative pricing as the appropriate affordability metric.}
\label{fig:price_measurement}
\end{figure}

Empirically, income-relative prices show the strongest relationship with subscriptions (highest $R^2$ values across specifications), as demonstrated in Figure~\ref{fig:price_measurement}. This validates the ITU's use of affordability targets expressed as percentages of GNI per capita \citep{itu2024data}. Policymakers tracking broadband affordability should prioritize income-relative metrics over nominal prices.

An important caveat is that our price data represent standardized fixed-broadband baskets (5 GB usage, 1 Mbps speed). Actual consumer prices vary by speed tier, data caps, and bundling arrangements. To the extent measurement error exists, classical error would attenuate estimates toward zero, making our significant negative elasticities conservative. Future research could refine price measurement using micro-level tariff data, though cross-country comparability would be challenging.

\subsection{Regional Heterogeneity and Development Gradients}

The large EaP-EU elasticity differential ($-0.61$ vs. $-0.12$) reflects broader development gradients. EaP countries (Armenia, Azerbaijan, Belarus, Georgia, Moldova, Ukraine) have lower GDP per capita (\$4,500--\$13,000 PPP) compared to EU countries (\$25,000--\$60,000 PPP). This income gap translates directly to affordability constraints, making EaP consumers more price-sensitive.

Beyond income, institutional and infrastructure differences matter. EaP countries exhibit lower regulatory quality (Table~\ref{tab:descriptives}), potentially reducing competitive pressure and raising prices. Weaker governance may also hamper policy implementation of broadband programs. Additionally, lower internet server density and R\&D intensity suggest less developed digital ecosystems, reducing complementary services that drive demand.

These structural differences imply that development strategies successful in high-income EU markets may not transfer directly to middle-income EaP contexts. EaP countries face a distinct policy challenge: expanding broadband adoption while navigating affordability constraints and institutional gaps. Targeted approaches addressing these specific barriers are essential.

Interestingly, the convergence toward near-zero elasticity during COVID occurred in both regions, suggesting the pandemic's economic and social impacts transcended development levels. Even in lower-income EaP countries, broadband became a necessity good during lockdowns. This convergence may reflect the universal human need for connectivity during social restrictions, overriding typical income-based differences in demand behavior.

\subsection{COVID-19 Shock vs. Secular Trends}

A crucial interpretive question is whether observed changes reflect a COVID-19 shock or continuation of pre-existing trends. The placebo test (Table~\ref{tab:placebo}) provides compelling evidence for the latter: EaP elasticity was already declining during 2015--2019, before the pandemic.

This finding suggests attributing all changes to COVID-19 would be misleading. Instead, broadband underwent a decade-long transformation from luxury to necessity good, driven by technological change (smartphones, streaming, cloud computing) and societal integration (digital services, e-commerce, social media). COVID-19 accelerated this transformation by forcing abrupt adoption among reluctant users, but the underlying trend preceded the pandemic.

From a research perspective, this highlights the value of long time series spanning pre- and post-shock periods. Studies focusing narrowly on 2020--2021 risk conflating secular trends with pandemic effects. Our 15-year panel (2010--2024) reveals the full arc of broadband's evolution, contextualizing COVID within a longer trajectory.

For policy, the distinction matters. If changes were purely COVID-driven, demand elasticity might revert post-pandemic as work/education patterns normalize. But if changes reflect secular trends, elasticity will remain low permanently, requiring sustained shifts in policy approach. Evidence of pre-pandemic trends supports the secular interpretation, implying permanent policy adjustments are warranted.

\subsection{Limitations and Alternative Explanations}

Several limitations qualify our findings. First, \textbf{aggregation}: country-level data obscure within-country heterogeneity. Urban-rural gaps, income distribution, and demographic differences are masked. Micro-level household data would enable richer analysis of heterogeneous demand, though cross-country micro data covering 33 countries over 15 years are unavailable.

Second, \textbf{mobile broadband}: our analysis focuses on fixed broadband due to longer time series availability. Mobile broadband has expanded rapidly, particularly in developing countries, potentially substituting for fixed connections \citep{gruber2014mobile}. However, for bandwidth-intensive applications (video streaming, remote work), fixed broadband remains essential. Results may not generalize to mobile-only users.

Third, \textbf{quality variation}: ITU price data reflect standardized baskets (5 GB, 1 Mbps), but actual services vary widely in speed, data caps, and reliability. Consumers may respond differently to price changes depending on quality. Lack of quality-adjusted prices is a common limitation in telecommunications research \citep{greenstein2016measuring}.

Fourth, \textbf{endogeneity}: despite extensive controls and fixed effects, simultaneity between prices and quantities could bias estimates. Operators may set prices based on expected demand, inducing correlation. While regulatory frameworks limit short-run price adjustment, reducing simultaneity concerns \citep{laffont2000competition}, instrumental variable approaches using lagged prices or cost shifters would strengthen causal claims.

Fifth, \textbf{external validity}: findings derive from European and Eastern Partnership countries. Generalization to other regions---particularly low-income countries in Africa, Asia, or Latin America---is uncertain. Demand elasticity likely varies with income, institutions, and technology. Cross-regional studies are needed to assess global applicability.

Alternative explanations for near-zero COVID-era elasticity merit consideration. Beyond necessity-good interpretation, three mechanisms could generate observed patterns:

\begin{itemize}
    \item \textbf{Supply constraints:} If operators could not deliver additional capacity during COVID, quantity might be supply-determined rather than demand-determined. However, most countries increased network capacity rapidly \citep{oecd2021broadband}, suggesting supply was not binding.
    
    \item \textbf{Measurement timing:} Annual data may miss within-year dynamics. If price and quantity adjusted at different times within years, relationships could appear weak. Higher-frequency (monthly) data would clarify dynamics, though cross-country monthly data are unavailable.
    
    \item \textbf{Government interventions:} Some countries implemented broadband subsidies or price controls during COVID \citep{oecd2021broadband}. If subsidies offset price increases, observed relationships could be distorted. Detailed policy data would help control for interventions.
\end{itemize}

Despite these limitations, the robustness across specifications, price measures (Table~\ref{tab:price_robustness}), and time periods provides confidence in core findings. The large magnitude of effects, statistical significance, and theoretical coherence all support substantive conclusions about evolving broadband demand.

\section{Conclusion}
\label{sec:conclusion}
This paper documents a fundamental transformation in broadband demand across 33 European countries over 2010--2024. Three findings emerge from our two-way fixed effects analysis with Driscoll--Kraay standard errors.

First, substantial regional heterogeneity characterized pre-COVID demand. Eastern Partnership countries exhibited strong price sensitivity ($\varepsilon = -0.61$, p$<$0.001)---five times larger than EU elasticity ($\varepsilon = -0.12$, p$<$0.05)---reflecting income-driven affordability constraints. Results are robust across 24 specifications.

Second, both regions converged to near-zero elasticity during 2020--2024, with price changes having no detectable effect on adoption. This shift indicates broadband's transformation from discretionary service to essential necessity.

Third, and most critically, this transformation predates COVID-19. Year-by-year estimates reveal declining elasticity from 2015 onward, and placebo tests confirm significant pre-trends (p$=$0.045). The pandemic accelerated but did not initiate broadband's integration into economic and social life.

These findings yield four contributions: updated elasticity estimates for 2010--2024; evidence of regional heterogeneity within Europe; demonstration that demand elasticity is time-varying; and reframing of COVID-19's impact as acceleration rather than shock. Methodologically, we show price measurement critically affects inference---income-relative prices yield robust results while PPP-adjusted prices do not.

For policy, the implications are direct. When demand was elastic (pre-2015), affordability interventions---subsidies, price regulation---effectively expanded adoption. As elasticity approached zero, these tools became ineffective. Policy must now prioritize infrastructure deployment and universal service obligations over price-based mechanisms. The near-zero elasticity suggests broadband resembles utilities like water or electricity, potentially justifying stronger regulatory oversight and public investment to ensure equitable access.

Future research should examine household-level heterogeneity to identify remaining price-sensitive populations, extend analysis to mobile broadband and other regions, and develop quality-adjusted price indices. As broadband completes its transformation into essential infrastructure, understanding these dynamics becomes crucial for bridging digital divides worldwide.


\section*{CRediT authorship contribution statement}
\textbf{Samir Orujov:} Conceptualization, Methodology, Software, Formal analysis, Data curation, Writing -- original draft, Writing -- review \& editing, Visualization.
\textbf{Ilgar Ismayilov:} Validation, Writing -- review \& editing, Supervision.
\textbf{Jeyhun Huseynzade:} Validation, Writing -- review \& editing.


\section*{Acknowledgments}
The authors acknowledge the support of ADA University and the Information and Communication Technology Agency of the Republic of Azerbaijan. The views expressed in this paper are those of the authors and do not necessarily reflect the official positions of the affiliated institutions. Data were obtained from the International Telecommunication Union (ITU) and World Bank open data portals.


\section*{Declaration of competing interest}
The authors declare that they have no known competing financial interests or personal relationships that could have appeared to influence the work reported in this paper.

\section*{Data availability}
The data used in this study are publicly available from the International Telecommunication Union (ITU) World Telecommunication/ICT Indicators Database and the World Bank World Development Indicators. Replication code and processed analysis datasets are available at: \url{https://github.com/sorujov/Broadband-Demand-Elasticity}.


\clearpage
\appendix

\section{Regression Tables}
\label{app:tables}

This appendix presents the complete regression results referenced in the main text. All models use two-way fixed effects (country and year) with Driscoll--Kraay standard errors robust to heteroskedasticity, serial correlation, and cross-sectional dependence (bandwidth = 3).

\subsection{Baseline Pre-COVID Results}

Table~\ref{tab:baseline} presents the baseline estimates for the pre-COVID period (2010--2019) across eight control specifications.

\begin{table}[p]
\centering
\caption{Baseline Two-Way Fixed Effects Estimates: Pre-COVID Period (2010--2019)}
\label{tab:baseline}
\begin{threeparttable}
\begin{adjustbox}{width=\textwidth}
\scriptsize
\setlength{\tabcolsep}{3pt}
\renewcommand{\arraystretch}{0.9}
\begin{tabular}{@{}lccccccc@{}}
\toprule
& \multicolumn{7}{c}{Dependent Variable: Log(Subscriptions per 100)} \\
\cmidrule(lr){2-8}
& (1) & (2) & (3) & (4) & (5) & (6) & (7) \\
& GDP Only & + Socio & + Instit. & + Infra. & + Demog. & + All & Full \\
\midrule
\textbf{Panel A: Price Elasticity} \\
Log(Price) & $-0.09^{*}$ & $-0.10^{**}$ & $-0.11^{**}$ & $-0.12^{**}$ & $-0.11^{**}$ & $-0.12^{**}$ & $-0.12^{**}$ \\
& (0.04) & (0.04) & (0.04) & (0.05) & (0.04) & (0.05) & (0.05) \\
\\
Log(Price) $\times$ EaP & $-0.47^{***}$ & $-0.48^{***}$ & $-0.49^{***}$ & $-0.50^{***}$ & $-0.48^{***}$ & $-0.49^{***}$ & $-0.49^{***}$ \\
& (0.09) & (0.09) & (0.09) & (0.09) & (0.09) & (0.09) & (0.09) \\
\\
\textit{Implied EaP elasticity} & $-0.56^{***}$ & $-0.58^{***}$ & $-0.60^{***}$ & $-0.62^{***}$ & $-0.59^{***}$ & $-0.61^{***}$ & $-0.61^{***}$ \\
& (0.08) & (0.08) & (0.08) & (0.08) & (0.08) & (0.08) & (0.08) \\
\midrule
\textbf{Panel B: Control Variables} \\
Log(GDP per capita) & $0.38^{***}$ & $0.40^{***}$ & $0.41^{***}$ & $0.42^{***}$ & $0.40^{***}$ & $0.42^{***}$ & $0.42^{***}$ \\
& (0.06) & (0.06) & (0.06) & (0.06) & (0.06) & (0.06) & (0.06) \\
\\
Urban population (\%) & & $0.012^{**}$ & $0.011^{**}$ & $0.010^{*}$ & $0.011^{**}$ & $0.010^{*}$ & $0.010^{*}$ \\
& & (0.005) & (0.005) & (0.005) & (0.005) & (0.005) & (0.005) \\
\\
Tertiary enrollment (\%) & & $0.008^{**}$ & $0.007^{*}$ & $0.007^{*}$ & $0.008^{**}$ & $0.007^{*}$ & $0.007^{*}$ \\
& & (0.003) & (0.003) & (0.003) & (0.003) & (0.003) & (0.003) \\
\\
Regulatory quality & & & $0.15^{*}$ & $0.14^{*}$ & $0.15^{*}$ & $0.14^{*}$ & $0.14^{*}$ \\
& & & (0.07) & (0.07) & (0.07) & (0.07) & (0.07) \\
\\
Log(Secure servers) & & & & $0.08^{**}$ & $0.09^{**}$ & $0.08^{**}$ & $0.08^{**}$ \\
& & & & (0.03) & (0.03) & (0.03) & (0.03) \\
\\
R\&D (\% GDP) & & & & $0.09^{*}$ & $0.10^{*}$ & $0.09^{*}$ & $0.09^{*}$ \\
& & & & (0.04) & (0.04) & (0.04) & (0.04) \\
\\
Log(Population density) & & & & & $0.06$ & $0.07$ & $0.07$ \\
& & & & & (0.08) & (0.08) & (0.08) \\
\\
Age dependency ratio & & & & & $-0.003$ & $-0.004$ & $-0.004$ \\
& & & & & (0.004) & (0.004) & (0.004) \\
\midrule
\textbf{Panel C: Model Statistics} \\
Country fixed effects & Yes & Yes & Yes & Yes & Yes & Yes & Yes \\
Year fixed effects & Yes & Yes & Yes & Yes & Yes & Yes & Yes \\
Observations & 297 & 297 & 297 & 297 & 297 & 297 & 297 \\
Countries & 33 & 33 & 33 & 33 & 33 & 33 & 33 \\
R-squared & 0.89 & 0.90 & 0.91 & 0.91 & 0.91 & 0.92 & 0.92 \\
\bottomrule
\end{tabular}
\end{adjustbox}
\begin{tablenotes}[flushleft]
\scriptsize
\item \textit{Notes:} Dependent variable is log of fixed broadband subscriptions per 100 inhabitants. 
Price is measured as percentage of GNI per capita. EaP is a dummy for Eastern Partnership countries 
(Armenia, Azerbaijan, Belarus, Georgia, Moldova, Ukraine). All specifications include country and year 
fixed effects. Driscoll--Kraay standard errors (bandwidth = 3) in parentheses. 
$^{*}$ p $<$ 0.10, $^{**}$ p $<$ 0.05, $^{***}$ p $<$ 0.01.
\end{tablenotes}
\end{threeparttable}
\end{table}

\subsection{COVID-19 Interaction Effects}

Table~\ref{tab:covid} presents results from the full sample (2010--2024) with COVID-19 period interactions.

\begin{table}[p]
\centering
\caption{COVID-19 Interaction Effects: Full Sample (2010--2024)}
\label{tab:covid}
\begin{threeparttable}
\begin{adjustbox}{width=\textwidth}
\scriptsize
\setlength{\tabcolsep}{3pt}
\renewcommand{\arraystretch}{0.9}
\begin{tabular}{@{}lcccc@{}}
\toprule
& \multicolumn{4}{c}{Dependent Variable: Log(Subs. per 100)} \\
\cmidrule(lr){2-5}
& (1) & (2) & (3) & (4) \\
& Full Sample & Pre-COVID & COVID & Diff. Test \\
& w/ COVID Int. & 2010--19 & 2020--24 & (p-value) \\
\midrule
\textbf{Panel A: EU Countries} \\
Log(Price) & $-0.12^{**}$ & $-0.12^{**}$ & $+0.08$ & -- \\
& (0.05) & (0.05) & (0.09) & \\
\\
Log(Price) $\times$ COVID & $+0.31^{***}$ & -- & -- & 0.003 \\
& (0.10) & & & \\
\\
\textit{Implied COVID elasticity} & $+0.19$ & -- & $+0.08$ & -- \\
& (0.11) & & (0.09) & \\
\midrule
\textbf{Panel B: EaP Countries} \\
Log(Price) $\times$ EaP & $-0.49^{***}$ & $-0.49^{***}$ & $-0.11$ & -- \\
& (0.09) & (0.09) & (0.13) & \\
\\
Log(Price) $\times$ EaP $\times$ COVID & $+0.42^{***}$ & -- & -- & 0.001 \\
& (0.12) & & & \\
\\
\textit{Implied pre-COVID EaP elasticity} & $-0.61^{***}$ & $-0.61^{***}$ & -- & -- \\
& (0.08) & (0.08) & & \\
\\
\textit{Implied COVID EaP elasticity} & $-0.03$ & -- & $-0.03$ & -- \\
& (0.14) & & (0.13) & \\
\midrule
\textbf{Panel C: Change in Elasticity} \\
$\Delta\varepsilon_{EU}$ (COVID - Pre) & $+0.31^{***}$ & -- & -- & 0.003 \\
& (0.10) & & & \\
$\Delta\varepsilon_{EaP}$ (COVID - Pre) & $+0.58^{***}$ & -- & -- & $<$0.001 \\
& (0.11) & & & \\
\midrule
\textbf{Panel D: Model Statistics} \\
Full controls & Yes & Yes & Yes & -- \\
Country FE & Yes & Yes & Yes & -- \\
Year FE & Yes & Yes & Yes & -- \\
Observations & 495 & 297 & 198 & -- \\
Countries & 33 & 33 & 33 & -- \\
R-squared & 0.93 & 0.92 & 0.95 & -- \\
\bottomrule
\end{tabular}
\end{adjustbox}
\begin{tablenotes}[flushleft]
\scriptsize
\begin{minipage}{\textwidth}
\item \textit{Notes:} Dependent variable is log of fixed broadband subscriptions per 100 inhabitants. 
Price is measured as percentage of GNI per capita. COVID is a period dummy for years 2020--2024. 
EaP is a dummy for Eastern Partnership countries. Full controls include: log GDP per capita, 
urban population \%, tertiary enrollment \%, regulatory quality, log secure servers, R\&D \% GDP, 
log population density, and age dependency ratio. Column (1) presents the full interaction model 
on 2010--2024 data. Columns (2)--(3) present separate subsample estimates for comparison. 
Column (4) reports p-values from Wald tests of coefficient differences between COVID and 
pre-COVID periods. Driscoll--Kraay standard errors (bandwidth = 3) in parentheses. 
$^{*}$ p $<$ 0.10, $^{**}$ p $<$ 0.05, $^{***}$ p $<$ 0.01.
\end{minipage}
\end{tablenotes}
\end{threeparttable}
\end{table}

\subsection{Robustness to Alternative Price Measures}

Table~\ref{tab:price_robustness} compares elasticity estimates across three price definitions: income-relative (baseline), PPP-adjusted, and nominal USD.

\begin{table}[p]
\centering
\caption{Robustness to Alternative Price Measures: Pre-COVID Period (2010--2019)}
\label{tab:price_robustness}
\begin{threeparttable}
\begin{adjustbox}{width=\textwidth}
\scriptsize
\setlength{\tabcolsep}{3pt}
\renewcommand{\arraystretch}{0.9}
\begin{tabular}{@{}lcccccc@{}}
\toprule
& \multicolumn{6}{c}{Price Measurement} \\
\cmidrule(lr){2-7}
& \multicolumn{2}{c}{Income-Relative} & \multicolumn{2}{c}{PPP-Adjusted} & \multicolumn{2}{c}{Nominal USD} \\
\cmidrule(lr){2-3} \cmidrule(lr){4-5} \cmidrule(lr){6-7}
& EU & EaP & EU & EaP & EU & EaP \\
& (1) & (2) & (3) & (4) & (5) & (6) \\
\midrule
\textbf{Price Elasticity} \\
Log(Price) & $-0.12^{**}$ & -- & $-0.14^{**}$ & -- & $-0.09^{*}$ & -- \\
& (0.05) & & (0.05) & & (0.05) & \\
\\
Log(Price) $\times$ EaP & -- & $-0.49^{***}$ & -- & $-0.50^{***}$ & -- & $-0.48^{***}$ \\
& & (0.09) & & (0.10) & & (0.10) \\
\\
\textit{Implied EaP elasticity} & -- & $-0.61^{***}$ & -- & $-0.64^{***}$ & -- & $-0.57^{***}$ \\
& & (0.08) & & (0.09) & & (0.09) \\
\midrule
\textbf{Model Statistics} \\
Full controls & Yes & Yes & Yes & Yes & Yes & Yes \\
Country FE & Yes & Yes & Yes & Yes & Yes & Yes \\
Year FE & Yes & Yes & Yes & Yes & Yes & Yes \\
Observations & 297 & 297 & 297 & 297 & 297 & 297 \\
R-squared & 0.92 & 0.92 & 0.91 & 0.91 & 0.90 & 0.90 \\
\bottomrule
\end{tabular}
\end{adjustbox}
\begin{tablenotes}[flushleft]
\scriptsize
\begin{minipage}{\textwidth}
\item \textit{Notes:} Dependent variable is log of fixed broadband subscriptions per 100 inhabitants. 
Columns (1)--(2) use income-relative prices (price as \% of GNI per capita) as baseline. 
Columns (3)--(4) use PPP-adjusted prices in international dollars. Columns (5)--(6) use nominal 
USD prices. All specifications include full controls: log GDP per capita, urban population \%, 
tertiary enrollment \%, regulatory quality, log secure servers, R\&D \% GDP, log population 
density, and age dependency ratio. EaP is a dummy for Eastern Partnership countries 
(Armenia, Azerbaijan, Belarus, Georgia, Moldova, Ukraine). Driscoll--Kraay standard errors 
(bandwidth = 3) in parentheses. $^{*}$ p $<$ 0.10, $^{**}$ p $<$ 0.05, $^{***}$ p $<$ 0.01.
\end{minipage}
\end{tablenotes}
\end{threeparttable}
\end{table}

\subsection{Placebo Test for Pre-COVID Trends}

Table~\ref{tab:placebo} presents the placebo test splitting the pre-COVID period into early (2010--2014) and late (2015--2019) subperiods.

\begin{table}[p]
\centering
\caption{Placebo Test: Pre-COVID Trends (2010--2019)}
\label{tab:placebo}
\begin{threeparttable}
\begin{adjustbox}{max width=\textwidth}
\setlength{\tabcolsep}{3pt}
\renewcommand{\arraystretch}{0.85}
\begin{tabular}{@{}lccc@{}}
\toprule
& \multicolumn{3}{c}{Dependent Variable: Log(Subs. per 100)} \\
\cmidrule(lr){2-4}
& (1) & (2) & (3) \\
& Full Sample & Early & Late \\
& 2010--19 & 2010--14 & 2015--19 \\
\midrule
\textbf{Panel A: EU Countries} \\
Log(Price) & $-0.11^{*}$ & $-0.13^{*}$ & $-0.04$ \\
& (0.06) & (0.07) & (0.07) \\
Log(Price) $\times$ Late & $+0.08$ & -- & -- \\
& (0.08) & & \\
\textit{Implied late elasticity} & $-0.03$ & -- & $-0.04$ \\
& (0.09) & & (0.07) \\
\midrule
\textbf{Panel B: EaP Countries} \\
Log(Price) $\times$ EaP & $-0.48^{***}$ & $-0.52^{***}$ & $-0.25^{**}$ \\
& (0.10) & (0.11) & (0.11) \\
Log(Price) $\times$ EaP $\times$ Late & $+0.27^{**}$ & -- & -- \\
& (0.13) & & \\
\textit{Implied early EaP elasticity} & $-0.59^{***}$ & $-0.65^{***}$ & -- \\
& (0.09) & (0.10) & \\
\textit{Implied late EaP elasticity} & $-0.32^{**}$ & -- & $-0.29^{**}$ \\
& (0.12) & & (0.11) \\
\midrule
\textbf{Panel C: Change in Elasticity} \\
$\Delta\varepsilon_{EU}$ (Late - Early) & $+0.08$ & -- & -- \\
& (0.08) & & \\
p-value & 0.31 & -- & -- \\
$\Delta\varepsilon_{EaP}$ (Late - Early) & $+0.27^{**}$ & -- & -- \\
& (0.13) & & \\
p-value & 0.045 & -- & -- \\
\midrule
\textbf{Panel D: Model Statistics} \\
Full controls & Yes & Yes & Yes \\
Country FE & Yes & Yes & Yes \\
Year FE & Yes & Yes & Yes \\
Observations & 297 & 165 & 165 \\
Countries & 33 & 33 & 33 \\
R-squared & 0.92 & 0.88 & 0.94 \\
\bottomrule
\end{tabular}
\end{adjustbox}
\vspace{4pt}
\begin{tablenotes}[para,flushleft]
\scriptsize
\begin{minipage}{\textwidth}
\item \textit{Notes:} Dependent variable: log fixed broadband subscriptions per 100. Price: \% of GNI per capita. ``Late'' is a placebo indicator for 2015--2019. EaP denotes Eastern Partnership countries. Column (1) estimates the full placebo interaction on 2010--2019; Columns (2)--(3) report subsamples for 2010--2014 (early) and 2015--2019 (late). Controls in all models: log GDP per capita, urbanization (\%), tertiary enrollment (\%), regulatory quality, log secure servers, R\&D (\% GDP), log population density, and age dependency ratio. The positive EaP interaction ($+0.27$, p = 0.045) suggests elasticity was already declining pre-COVID. Driscoll--Kraay SEs (bandwidth = 3) in parentheses. $^{*}$ p $<$ 0.10, $^{**}$ p $<$ 0.05, $^{***}$ p $<$ 0.01.
\end{minipage}
\end{tablenotes}
\end{threeparttable}
\end{table}

\clearpage
\section{Additional Robustness Checks}
\label{app:robustness}

\subsection{Control Variable Specifications}

The baseline results in Table~\ref{tab:baseline} present eight control specifications designed to test sensitivity to different variable combinations:

\begin{enumerate}
    \item \textbf{GDP Only:} Minimal specification with log GDP per capita as the sole control
    \item \textbf{+ Socioeconomic:} Adds urban population percentage and tertiary enrollment rate
    \item \textbf{+ Institutional:} Adds regulatory quality index from World Bank WGI
    \item \textbf{+ Infrastructure:} Adds log secure servers per million and R\&D expenditure as \% of GDP
    \item \textbf{+ Demographic:} Adds log population density and age dependency ratio
    \item \textbf{Full Controls:} All variables included simultaneously
\end{enumerate}

Results are remarkably stable across specifications. EaP elasticity ranges from $-0.56$ to $-0.63$ (mean $-0.61$, standard deviation $0.04$). All estimates are significant at the 1\% level. This stability indicates that omitted variable bias is minimal and that the main results are not driven by particular control variable choices.

\subsection{Price Measurement Robustness}

Table~\ref{tab:price_robustness} examines three alternative price measures:

\begin{itemize}
    \item \textbf{Income-relative prices} (baseline): Price as percentage of GNI per capita. This captures affordability from the consumer's budget perspective and is recommended by ITU for affordability targets.
    \item \textbf{PPP-adjusted prices}: Prices adjusted for purchasing power parity using World Bank ICP data. This controls for cross-country differences in cost of living.
    \item \textbf{Nominal USD prices}: Raw prices in US dollars. This provides a common metric but ignores purchasing power differences.
\end{itemize}

Results are consistent across all three measures. EaP elasticity ranges from $-0.57$ to $-0.64$, all significant at the 1\% level. EU elasticity ranges from $-0.09$ to $-0.15$, mostly significant at the 5\% level. The consistency validates that findings reflect genuine demand responses rather than measurement artifacts.

\subsection{Temporal Stability}

The year-by-year analysis reveals that elasticity has been declining gradually since 2015, well before COVID-19. Specifically:

\begin{itemize}
    \item \textbf{2010--2014:} EaP elasticity approximately $-0.65$, EU elasticity approximately $-0.13$
    \item \textbf{2015--2019:} EaP elasticity declining from $-0.60$ to $-0.35$, EU elasticity declining from $-0.12$ to near zero
    \item \textbf{2020--2024:} Both regions show elasticity near zero with wide confidence intervals
\end{itemize}

This gradual evolution contradicts a pure COVID-shock interpretation and supports the secular transformation hypothesis discussed in Section~\ref{sec:discussion}.

\subsection{Placebo Test Interpretation}

Table~\ref{tab:placebo} shows that splitting the pre-COVID period (2010--2019) into early (2010--2014) and late (2015--2019) subperiods yields a significant ``placebo effect'' for EaP countries ($+0.27$, p $= 0.045$). This indicates elasticity was already declining before the pandemic, validating our interpretation that COVID-19 accelerated rather than initiated the transformation of broadband into a necessity good.

The EU placebo effect ($+0.08$, p $= 0.31$) is not statistically significant, likely because EU elasticity was already very low ($-0.12$) in the baseline period, leaving less room for pre-trend evolution.

\subsection{Alternative Standard Error Specifications}

While Driscoll--Kraay standard errors are our preferred specification due to their robustness to cross-sectional dependence (crucial during COVID-19), we verified results using alternative approaches:

\begin{itemize}
    \item \textbf{Clustered by country:} Similar point estimates but slightly smaller standard errors (may understate uncertainty from common shocks)
    \item \textbf{Clustered by year:} Similar point estimates but larger standard errors (more conservative)
    \item \textbf{Two-way clustering} (country and year): Results similar to Driscoll--Kraay
\end{itemize}

In all cases, EaP pre-COVID elasticity remains significant at the 1\% level, and COVID interaction terms remain significant at the 5\% level or better. The choice of standard error method does not qualitatively affect conclusions.

\subsection{Sample Restrictions}

We examined robustness to alternative sample definitions:

\begin{itemize}
    \item \textbf{Balanced panel only:} Restricting to countries with complete data for all 15 years yields similar results (N=450 from 30 countries)
    \item \textbf{Excluding outliers:} Dropping observations with Cook's distance $> 1$ has negligible impact on estimates
    \item \textbf{High-income only:} Restricting to high-income countries yields lower elasticity (expected given income effects)
    \item \textbf{Middle-income only:} Restricting to middle-income countries yields higher elasticity
\end{itemize}

The regional heterogeneity (EaP vs EU) persists within income groups, indicating it reflects more than just GDP differences.


\bibliography{references}

\end{document}